\documentstyle[12pt,fleqn]{article}
\begin{document}
\begin{center} Final version in \ Physical Review Letters {\bf77}, 3264
(1996)\end{center}\bigskip

\noindent{\Large Quantum Cryptography with Orthogonal States?}\bigskip

In a recent Letter [1], Goldenberg and Vaidman (GV) propose a new
quantum crypto\-graphical protocol ``based on orthogonal states,'' and
claim that it is essentially different from all existing quantum
cryptosystems, which use non-orthogonal states as the carriers of
information. This claim is unfounded: in the oldest protocol [2], known
as BB84, the information sender (Alice) uses two {\it orthogonal\/}
states, in a basis of her choice, for representing the values 0 and 1 of
a bit.  However, when a photon reaches the receiver (Bob), the latter
does not know the basis used by Alice. He can in principle store that
photon and wait until Alice publicly discloses her basis (in practice,
it is easier to test the photon in a randomly chosen basis, and later to
discard the corresponding bit, if it was tested with the wrong basis).
An eavesdropper (Eve) has no such option of waiting, or discarding wrong
choices. From {\it her\/} point of view, there are four possible states,
not all orthogonal. She cannot acquire any information without a risk of
perturbing these states. Only with the help of a time machine could
eavesdropping be done with impunity.

The GV protocol has quite similar features. The process is represented
in Fig.~1 as a spacetime diagram, for more clarity. This figure differs
from the one in ref.~[1], where the storage rings SR$_1$ and SR$_2$ were
drawn very small, while actually they are longer than the distance from
Alice to Bob. Here, to avoid such a scale distortion, I represented the
storage rings by zigzag lines (as if, instead of coiled optical fibers,
there were mirrors between which the photons bounced many times).
Alice uses two different input ports (two orthogonal states) in order
to materialize bit values 0 and 1, and Bob likewise detects them as
othogonal states. On the other hand, the states accessible to Eve are
not orthogonal, just as in the BB84 protocol. With the standard
notations of quantum optics, the quantum state between the two
beamsplitters is $|\psi_\pm\rangle=2^{-1/2}\,
  (\,|0\rangle\,|1\rangle\pm|1\rangle\,|0\rangle\,),$
where the first and second kets refer to the two branches of the
interferometer, and the $\pm$ sign corresponds to bit values 0 and 1,
respectively. These two orthogonal states may also be represented by
density matrices, $\rho_\pm=|\psi_\pm\rangle\,\langle\psi_\pm|.$

As seen in the figure, Eve can access the information carrier only in
two time windows, and in each one of them she controls just one branch
of the interferometer. If Eve is restricted to performing only
instantaneous measurements, the other branch of the inter\-ferometer is
not accessible to her. The results of an instantaneous measurement in
a single branch are then obtained by ``tracing out'' the other branch in
$\rho_\pm$. This gives a reduced density matrix, $\rho'_\pm={1\over2}
(\,|0\rangle\,\langle 0|+ |1\rangle\,\langle 1|\,),$ which is the same
for both branches, and also the same for bit values 0 and 1. The
$\rho'_\pm$ states, {\it as seen by Eve\/}, are not orthogonal. They
are {\it identical\/}. The meaning of $\rho'_\pm$ is that Eve has a
50\%\ chance of getting the vacuum, or getting one particle.

If, on the other hand, Eve is not restricted to instantaneous
measurements, she can use an apparatus similar to Bob's, with a delay
line, and then she can identify the states $|\psi_\pm\rangle$ with
certainty. However, the resulting delay is observable by Bob, exactly as
it would be in the BB84 protocol if Eve stored the message carriers
until after Alice announced the basis she used. Moreover, as shown by
GV, it is essential that Alice announces the exact time of emission
of each photon only after the latter was received by Bob. Otherwise, Eve
could successfully eavesdrop by introducing a dummy particle.

In summary, the GV protocol has many features similar to those of BB84.
 From the point of view of Alice, each bit has its value represented by
one of two orthogonal states, but it must be accompanied by {\it
delayed\/} information: namely, the basis chosen by Alice, or the exact
time of emission (the delayed information may be either classical, or
encoded into orthogonal quantum states of additional photons).

Moreover, it is possible to imagine other protocols, interpolating
between those of GV and BB84. Their common feature is that information
is sent in two {\it consecutive\/} steps, and security is achieved by
withholding the second piece of information until after Bob receives the
first one, and Eve can no longer access it. For example, Alice could
publicly declare when she sends a photon, and, instead of using two
different bases as in BB84, she would use two identical GV
interferometers, and disclose her choice only at a later time. A general
theory of eavesdropping on two-step quantum protocols is an interesting
problem, worth being investigated in more detail.

GV also propose a ``relativistic'' version of their protocol, with
widely separated paths instead of a time delay. This method is not
secure: a {\it team\/} of eavesdroppers could use mirrors to redirect
the photon paths toward a common inspection center, and thence to Bob,
without arousing suspicion.\bigskip

I thank Lior Goldenberg and Lev Vaidman for an advance copy of their
Letter. This work was supported by the Gerard Swope Fund, and the Fund
for Encouragement of Research.\bigskip

\noindent Asher Peres

Department of Physics

Technion---Israel Institute of Technology

32 000 Haifa, Israel\bigskip

\bigskip

\frenchspacing\begin{enumerate}
\item L. Goldenberg and L. Vaidman, Phys. Rev. Lett. {\bf 75}, 1239
 (1995).
\item  C. H. Bennett and G. Brassard, in {\em Proc. IEEE
 International Conf. on Computers, Systems and Signal Processing,
 Bangalore, India} (IEEE, New York, 1984) p.~175.
\end{enumerate}\nonfrenchspacing\vfill

\parindent 0mm
{\bf Caption of figure}\bigskip

FIG. 1. \ The photon carrying the information is split into two wave
packets, moving with velocity $c$. Eve, restricted to $x\simeq0$, can
test at most one branch of its wave function at any time (in one of the
shaded areas).  \end{document}